\documentclass[preprint]{aastex}

\newcommand\gsim{\mathrel{\hbox{\rlap{\hbox{\lower4pt\hbox{$\sim$}}}\hbox{$>$}}}}

\usepackage[english]{babel}

\shorttitle{Galaxy Luminosity Profiles}
\shortauthors{Coenda et al.}

\begin{document}

\title{Photometric observations of SARS clusters: Galaxy Luminosity Profiles}

\author{
Valeria Coenda\altaffilmark{1},
Carlos Jos\'e Donzelli\altaffilmark{1},
Hernan Muriel\altaffilmark{1}}
\affil{Grupo de Investigaciones en Astronom\'\i a Te\'orica y Experimental,
IATE, Observatorio Astron\'omico, Universidad Nacional de
C\'ordoba, Laprida 854, X5000BGR, C\'ordoba, Argentina.}
\email{vcoenda@oac.uncor.edu, charly@oac.uncor.edu, hernan@oac.uncor.edu}
\author{Hernan Quintana,
Leopoldo Infante}
\affil{Departamento de Astronom\'\i a y Astrof\'\i sica,
Pontificia Universidad Cat\'olica, Vicu\~na Mackenna 4860, Casilla 306
Santiago 22, Chile.}
\email{hquintana@astro.puc.cl, linfante@astro.puc.cl}
\and
\author{Diego Garc\'{\i}a Lambas\altaffilmark{1}}
\affil{Grupo de Investigaciones en Astronom\'\i a Te\'orica y Experimental,
IATE, Observatorio Astron\'omico, Universidad Nacional de
C\'ordoba, Laprida 854, X5000BGR, C\'ordoba, Argentina.}
\email{dgl@oac.uncor.edu}
\altaffiltext{1}
{Consejo Nacional de Investigaciones Cient\'\i ficas y T\'ecnicas (CONICET),
Avenida Rivadavia 1917, C1033AAJ, Buenos Aires, Argentina.}

\begin{abstract}
We have analyzed CCD images of 14 Abell clusters in the R filter of the SARS survey (Way et al 2004), 
with $cz<40000kms^{-1}$. We have obtained the luminosity profiles for 507 galaxies of which 232 (46\%) have known redshifts. 
In order to fit the luminosity profiles we used the de Vaucouleurs law for bulge systems, an exponential profile for disk systems and we have also fitted the S\'{e}rsic's law $(r^n)$ to all galaxy profiles. We have found that 162 (32\%) galaxies in the sample have pure $r^{1/4}$ profiles, 168 (33\%) have pure exponential profiles, while 93 (18\%) galaxies have luminosity profiles that are well fitted by a combination of both bulge and disk profiles. On the other hand, we could not fit the classical bulge + disk profile to the remaining 84 (17\%) galaxies of the sample. For such cases we have only used the S\'ersic law. 
We have also analyzed how seeing and sky cleaning affect the structural and photometric parameters obtained through profile fitting. In addition, we have studied several 
relations between these parameters. We have found that bulges and disks show consistency with a unique relation in the $\mu_e - log(r_e)$ plane. We also found that bulges and disks obey a magnitude-size relation in the sense that large bulges and disks (large $r_e$ values) have high luminosities. On the other hand, S\'ersic law fitting parameters $n$, $rs$ and $\mu_s$ show a strong correlation in agreement with other authors suggesting the idea that not only ellipticals but all galaxies are likely to be understood as a one parameter family.
\end{abstract}


\section{Introduction}

Although it is widely accepted that the Hubble classification represents an underlying physical sequence, it is still difficult to construct a complete picture that describes the evolution of galaxy morphologies. To achieve a satisfactory understanding of the actual structure of galaxies and their evolution, it is important to separate the bulge and disk components.

Historically, different authors have used the $r^{1/4}$ law (de Vaucouleurs 1948) to model the light distribution of bulges and the exponential law (Freeman 1970) for disks. In recent years, astronomers have also been using the S\'ersic 
law (S\'ersic, 1968) to fit the surface brightness of galaxies. This new approach started with Davies et al. (1988) and was continued by Caon et al. (1993), Young \& Currie (1994), and was further discussed by Binggeli \& Jerjen (1997, 1998). Andredakis \& Sanders (1994), de Jong (1996), Carollo (1998) and others have found that the exponential profile provides a better fit to late type galaxies and suggested a range for the shape parameter $n$ (the exponent $n$ the S\'{e}rsic profile) to fit the bulge of spiral galaxies.

Structural parameters, such as the effective radius $r_e$, the effective surface brightness $\mu_{e}$ and the parameter $n$, are correlated by various relations, whose physical origins reside on the properties of the stellar populations of galaxies and on the dynamical structure of the systems. For the stellar population found in early type galaxies, one of the most interesting correlation between bulge structural parameters is the Kormendy relation (Kormendy 1977). This relation is a projection of the fundamental plane and it has the advantage that it can be constructed from photometric observations alone.  The slope of the Kormendy relation is an important tool to study the properties of the galaxy stellar population as a function of galaxy size (Ziegler et al. 1999). A similar relation was found for disk galaxies in clusters (Schade et al. 1996).

Several studies have been performed to determine the photometric properties of galaxy bulges and disks in clusters (Binggeli \& Jerjen 1997, 1998), and the possible correlation between these properties. More recently, Gavazzi et al. (2000) studied the near-infrared photometric properties of galaxies located in five nearby clusters. They found that less than 50\% of the elliptical galaxies have pure $r^{1/4}$ profiles. The majority of galaxies from E to Sb are best represented by $B+D$ profiles, while Scd galaxies have pure exponential profiles. In addition, they found that the type of decomposition depends on luminosity rather than on Hubble type. This is in agreement with the idea that ellipticals and dwarf ellipticals have different properties and therefore form two separate classes (Faber et al. 1997, Nieto et al. 1991). However, this idea stands in constrast with the findings of e.g. Jerjen \& Binggeli (1997), and others that there is only one scaling relation between the S\'ersic shape parameter and the absolute magnitude from giants to dwarfs over 8 orders of magnitudes, suggesting only one class of early-type galaxies. In a recent work Thomas (2002) studied the structural parameters of galaxies from the ENACS (ESO Nearby Abell Cluster Survey) and found correlations for structural and photometric parameters of the galaxy components.

In this paper we fit the luminosity profile and provide the photometric parameters for 507 galaxies selected from 14 southern Abell clusters of the SARS survey. We
also analyze the basic photometric scaling relations. The correlation 
between galaxy photometric parameters and cluster properties will be
the subject of a forthcoming paper. 

The paper is structured as follows: sample selection and observations are described in section \ref{data}, we examine the luminosity profiles of the cluster galaxies and we analyze the sky cleaning and seeing effects over the luminosity profiles in section \ref{method}. In section \ref{results} we show the profile fitting results and we examine the relations between the obtained photometric parameters. Finally, we summarize our conclusions in section \ref{concl}.

\section{Observations, data reduction and selected galaxies}\label{data}

We have obtained CCD images in the R Cousins filter of 14 Abell clusters with $cz <$ 40000 $km s^{-1}$ corresponding to the SARS survey (Southern Abell Clusters Redshifts Survey, Way et al. 2004).
The SARS  survey comprises Abell (1958) and Abell Corwin \& Olowin (1989) clusters with richness class $R\ge1$ in the region $-65^o\le \delta \le 0^o$ and $5^h\ge\alpha \le21^h$ (avoiding the LMC and SMC), with $b\le-40$. This redshift survey has more than 4000 galaxies that were selected from the APM catalog (Maddox et al. 1990) as follows: Galaxies brighter than $m_j=19$ and within 1.5x1.5 $deg^2$ centered on each cluster were pre-selected. Target galaxies were selected at random and the final completeness ($\sim$ 75\%) is roughly constant up to an apparent magnitude $\sim$ 18. Spectroscopic observations were carried out with the 2.5 meter DuPONT telescope of The Las Campanas Observatory, Chile.

The CCD images of the central region (21x21$\arcmin$) of the 14 SARS clusters analyzed in this paper were taken on August 1993 by two of us (HQ \& LI) with the Swope 1.0 m telescope at Las Campanas Observatory, Chile. A TEK CCD 2048$^2$ detector with a pixel scale of 0.61$\arcsec$ was used, covering a 20.8$\arcmin$ square area. In most cases we have taken for each cluster a set of 3 exposures of 900 s each, allowing us to clean-up the combined images for cosmic rays. During the run seeing conditions were good to fair, namely $FWHM$ = 1.4$\arcsec$ - 2.4$\arcsec$ and nights were photometric. In Table \ref{table1} we list the cluster sample, their central $\alpha$ and $\delta$ coordinates together with the log of observations and their radial velocity and radial velocity dispersion taken from Muriel et al. (2002).

Images were processed following the standard recipes for bias and flat-field corrections using IRAF routines. After this process sample images were carefully aligned using at least 7 stars selected from the borders and near to the center of the frames. Typical accuracy of the alignment was well under a pixel. Finally, aligned images were combined using a median algorithm in order to produce single R images.

Sky subtraction was performed by a polynomial fitting of first or second degree in specific sections of the images were no objects were present in order to avoid any kind of contamination (see Sect. \ref{Sky}).

The final step was the photometric calibration that was achieved using the Landolt UBVRI standard
star catalogue (Landolt 1992). An average of four Landolt star fields were observed each night 
at different airmasses, each field containing $\sim$ 5 stars which were used for the calibration. Error in zero point calibration was 0.06 mag. However, typical uncertainties of the galaxy photometry were in the range 0.06 - 0.1 mag. 
Fainter galaxies ($m \sim$ 16) have the largest errors and the main error source is background subtraction.

We have also fitted sky coordinates to the final R images using the plate solution
computed with the IRAF routine CCMAP. Field stars were identified using the SuperCOSMOS Sky Surveys. Average residuals for the astrometry were all under 1$\arcsec$.

In order to extract suitable galaxies for this work we have used SExtractor (Source Extractor). This routine is an automatic program that detects, classifies and performs photometry on sources from astronomical images (Bertin \& Arnouts 1996). Using SExtractor we have compiled the R 
magnitudes and positions of all galaxies from the whole cluster sample that have an apparent radius greater than 6$\arcsec$. The reason for this restriction is that we aimed to minimize seeing effects on the luminosity profile fitting results (see Sect. \ref{seeing effects}). This criteria for selecting galaxies by apparent diameter is more restrictive than the resulting limit in apparent magnitude and surface brightness. Consequently, we do not expect systematic effects associated to these parameters. The analyzed final sample consist in 507 galaxies of which 232 (46\%) have known redshifts (Way et al. 2004) and only 25 of these are not cluster members. For the remaining 275 (54\%) galaxies we have no redshift data available. 

\section{Luminosity profiles} \label{method}

Luminosity profiles were obtained using the $ellipse$ routine within STSDAS (Jedrezejewski 1987).
Profiles were fitted using the standard bulge + disk law:
\begin{equation} \label{dV+exp}
I(r)=I_{e}exp\Big[-7.688\Big[\Big(\frac{r}r_{e}\Big)^{1/4}-1\Big]\Big]+I_{0}exp\Big(-\frac{r}r_{0}\Big)
\end{equation}

In the above expression the first term corresponds to the bulge component (de Vaucouleurs 1948), being $I_{e}$ the intensity at $r_{e}$, the radius that encloses half of the total luminosity of the bulge (also known as the effective radius). The second term corresponds to the
disk component (Freeman, 1970), being $I_{0}$ the central intensity and $r_{0}$ the length scale. For some galaxies (10\% of the total sample) we noted that the profile was purely exponential plus a central gaussian excess. We have assumed for these galaxies that the source for the mentioned excess is an unresolved bulge and therefore we used for the profile fitting of this component the following expression:

\begin{equation} \label{gauss}
I(r)=I_{g}exp\Big(-2.71\Big(\frac{r}r_{g}\Big)^{2}\Big)
\end{equation}

In this case we were not able to calculate the bulge photometric parameters but its
total luminosity. 

On the other hand we have also used the S\'ersic law (S\'ersic 1968) to fit all galaxy
profiles:

\begin{equation} \label{sersic}
I(r)=I_{s}exp\Big(-\Big(\frac{r}r_{s}\Big)^n\Big)
\end{equation}

In this equation $I_{s}$ is the central intensity and $r_{s}$ the length scale. The exponent $n$ is a shape parameter that describes the amount of curvature in the profile. The case $n=0.25$ corresponds to the de Vaucouleurs law while $n=1$ corresponds to a disk profile. Equation \ref{sersic} is particularly useful since it needs only 3 parameters to fit the whole galaxy luminosity profile instead of the 4 required by expression \ref{dV+exp}. Moreover, several authors showed that S\'ersic's $n$ parameter correlate very well with the luminosity, the size (Caon, Capaccioli \& D'Onofrio 1993) and the bulge-to-disk ratio of the galaxies (Andredakis et al. 1995).

Total luminosities of both bulge and disk components were finally computed using the derived photometric parameters and integrating separately both terms of eq. \ref{dV+exp} as follows:

\begin{equation}
\label{luminosidad}
L=\int_{0}^{\infty}I(r)2\pi rdr
\end{equation}
which yields:
\begin{equation}
L_{bulge}=7.21\pi I_{e}r_{e}^{2}
\end{equation}
for the bulge component, and
\begin{equation}
L_{disk}=2\pi I_{0}r_{0}^{2}
\end{equation}
for the disk component. Note that eq. \ref{luminosidad} assumes that the galaxy is face on. Therefore, the intensity in the eq. \ref{dV+exp} and \ref{sersic} were corrected by inclination as Kent (1985). On the other hand, intensity (counts per pixel per second) was converted to surface brightness expressed in mag arcsec$^{-2}$ by the equation $\mu=-2.5log(I)$. Units of $r_e$, $r_0$ and $r_s$ are Kpc. Total apparent magnitudes were then converted into absolute magnitudes as follows: For member galaxies and galaxies with unknown redshifts we calculated their absolute magnitudes using the mean cluster redshift to avoid velocity dispersion uncertainties. For those not member galaxies we determined their absolute magnitudes using their own redshifts. Throughout this paper we have assumed a Hubble constant $H_0$ = 70 $km$ $s^{-1}$ $Mpc^{-1}$. 

\section{Profile fitting and parameter errors}

Parameters described in the above section  were obtained using the $nfit$ routine within
STSDAS (Schombert \& Bothun 1987). Fitting procedure is only reliable on the $S/N > 1$ portion of galaxy surface brightness profiles. Every error sources, like seeing, photon noise and sky cleaning, were carefully studied in order to check the uncertainties in the computed parameters. In our case, sky cleaning and seeing were the main error sources as well as galaxy overlapping. For the last case, we carefully masked the overlapped regions previous to the profile extraction. Then with the luminosity profiles and structural parameters obtained (center coordinates, ellipticity and position angle of the isophotes) we constructed a test galaxy that was subtracted to the original image and the resulting image was then used to extract the luminosity profile of the remaining galaxy. The process was repeated several times (2-3) until the profiles converged. 

\subsection{Sky cleaning effects}\label{Sky}

\begin{figure}
\begin{center}
\includegraphics[scale=0.6]{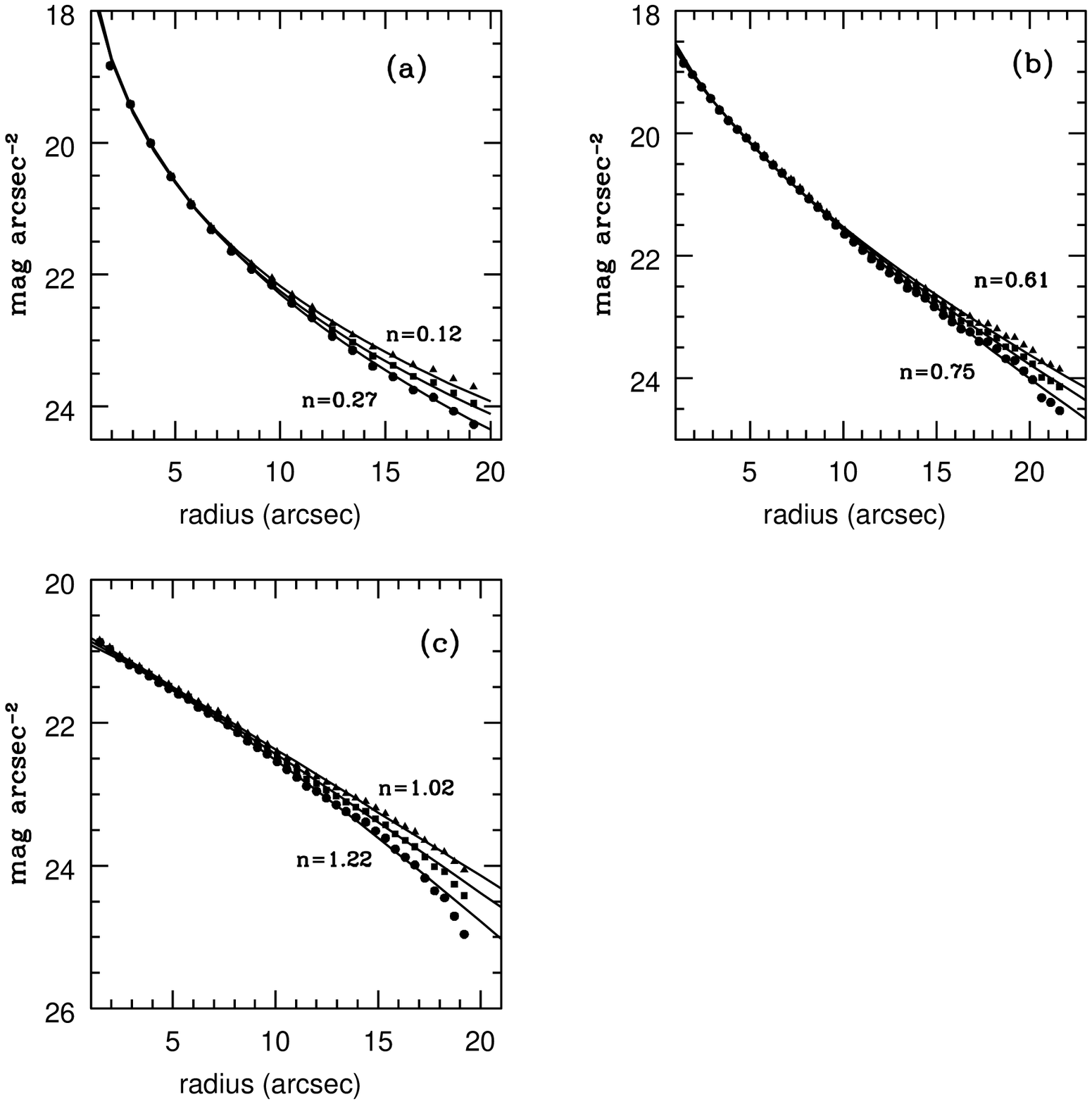}
\caption{Luminosity profiles for galaxies with different S\'ersic parameters.
Squares represent the original galaxy luminosity profile, while triangles and circles
represent the cases in which we artificially added and subtracted $\sigma_{sky}$ to the image, respectively.
(a) $n=0.18$, (b) $n=0.67$ and (c) $n=1.1$. }
\label{fig1}
\end{center}
\end{figure}

In most cases a two-dimensional first-degree polynomial was sufficient to give an accurate fit to the sky. The distribution of the residuals in the frame was used to estimate the uncertainty of the sky level, $\sigma_{sky}$ which has a deep influence on the faint end of the luminosity profiles and therefore on the computed fitted parameters. In order to quantify this effect we have created test images where we artificially added and subtracted a constant noise equals to $\sigma_{sky}$. We then extracted the new luminosity profiles 
as it was described in the previous section and we fitted them eq. \ref{dV+exp} and \ref{sersic}. This was done for several galaxies of different luminosities and sizes. Figure \ref{fig1} (a), (b) and (c) illustrates the case for the S\'ersic profile fitting. The galaxies have different luminosity profile shapes, $n=0.18$, $n=0.67$ and $n=1.10$. In general terms, errors in the parameters do not correlate with the shape of the luminosity profiles. Particularly, we found that the absolute error for the parameter $n$ is on average 0.08.

On the other hand, measured errors for $r_e$, $r_0$ and $r_s$ were never greater than 20\% while those calculated for $\mu_e$, $\mu_0$ and $\mu_s$ were below 0.25 mag arcsec$^{-2}$.

\subsection{Seeing effects} \label{seeing effects}

\begin{figure}
\begin{center}
\includegraphics[scale=0.6]{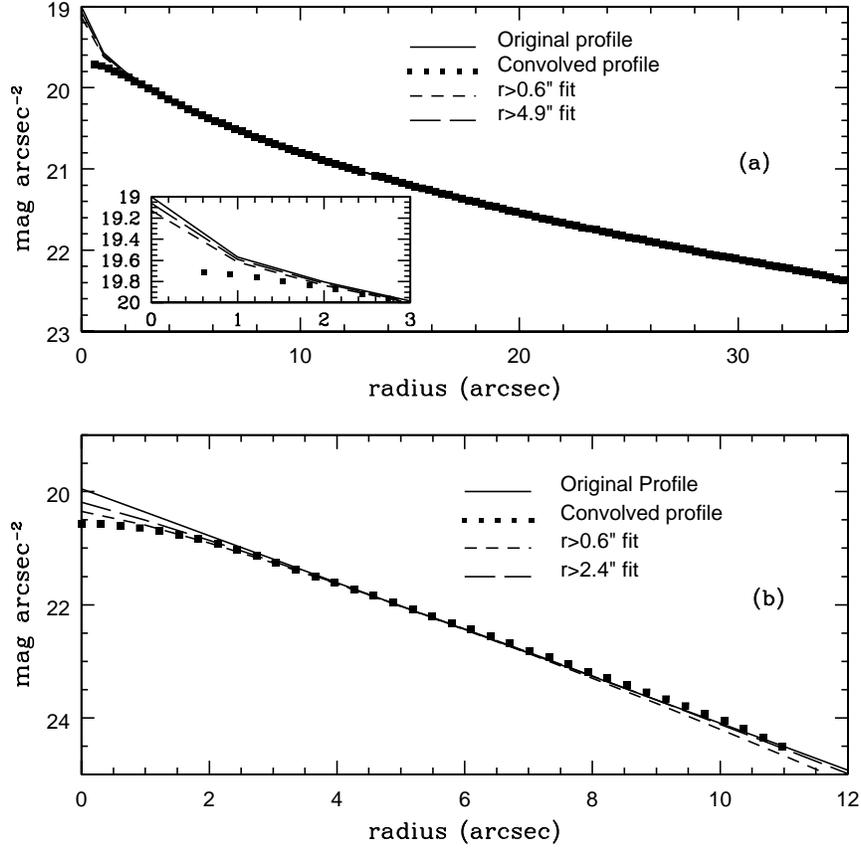}
\caption{(a) Galaxy 1: S\'{e}rsic law for a model galaxy with $\mu_{s}=19.44$ mag arcsec$^{-2}$, $r_{s}=3.7\arcsec$ and $n=0.50$, convolved with a FWHM gaussian filter of 2.4$\arcsec$ (squares). (b) Galaxy 2: S\'{e}rsic law for a model galaxy with $\mu_{s}=20.39$ mag arcsec$^{-2}$, $r_{s}=2.6\arcsec$ and $n=1.00$, convolved with a FWHM gaussian filter of 2.4$\arcsec$ (squares). Continuous line shows the original profile while short and long dashed lines show the obtained profiles by fitting the S\'ersic law beyond specific radius values.}
\label{fig2}
\end{center}
\end{figure}

There are many articles in which seeing and deconvolution effects on luminosity profiles are studied (Franx et al. 1989, Saglia et al. 1993, Trujillo et al. 2001). However, we decided to apply a very simple analysis that allowed us to determine, in terms of seeing, which is the most adequate region of the luminosity profile to perform the fitting procedure. In other words, we determined the minimum radius of the luminosity profile we can use for the fit in order to recover the 'true' photometric parameters of a galaxy. To achieve this goal we have generated model images of several galaxies with different photometric parameters, luminosities and sizes. These images were then convolved with a set of gaussian filters prior to the luminosity profile extraction. Finally we fitted eq. \ref{dV+exp} and \ref{sersic} to the obtained profiles in different semi-major axis regions. 
Two examples of our results are illustrated in Figure \ref{fig2} (a) and (b), and the obtained photometric parameters are listed in Table \ref{table2}. These cases consider two galaxies with different luminosities, $m_1$ = 13.65 (Galaxy 1) and $m_2$ = 16.50 (Galaxy 2), and different profile shapes $n = 0.5$ and $n = 1.0$. Selected values for the FWHM of the gaussian filters were chosen to simulate the seeing conditions of the observing run, 1.0 $\arcsec$ $\le$ FWHM $\le$ 2.4 $\arcsec$. We conclude that the more suitable region for profile fitting is that for $r$ beyond 1.5 FWHM. With this restriction, obtained parameters have associated errors that are well below those calculated in the previous section. Table \ref{table2} shows this effect. Note that for galaxy 1 (apparent radius at 1 $\sigma_{sky}$ detection $\sim$ 12 FWHM), obtained parameters are in reasonably agreement with the original ones even when we consider the whole luminosity profile for the fitting. However, this is not true for galaxy 2 that has an apparent radius $\sim$ 4.5 FWHM. In this case, the obtained parameters strongly depend on the interval considered and the best results are obtained by fitting data beyond 1.5 FWHM.

Most extreme cases are those for the galaxies with an apparent radius similar to 3 FWHM. For such cases, we could not recover the original photometric parameters since seeing mostly affects the shape of the whole luminosity profile. For this reason we decided to include in our analysis only galaxies with an apparent radius greater than 3$FWHM$ or $\sim 6\arcsec$.

\section{Results and Discussion}\label{results}

The derived photometric and structural parameters for the bulge and disk components are collected in table \ref{table3}. Galaxies are identified through their parent Abell number and their individual $\alpha$ and $\delta$ coordinates. In this table $\mu_e=-2.5 log(I_e)$, $\mu_0=-2.5 log(I_0)$ and $\mu_s=-2.5 log(I_s)$ expressed in mag arcsec$^{-2}$. Table \ref{table3} also lists the total absolute magnitudes for the galaxies as well as the total magnitudes for bulges and disks for those galaxies where profile decomposition was possible.

\begin{figure}
\begin{center}
\includegraphics[scale=0.5]{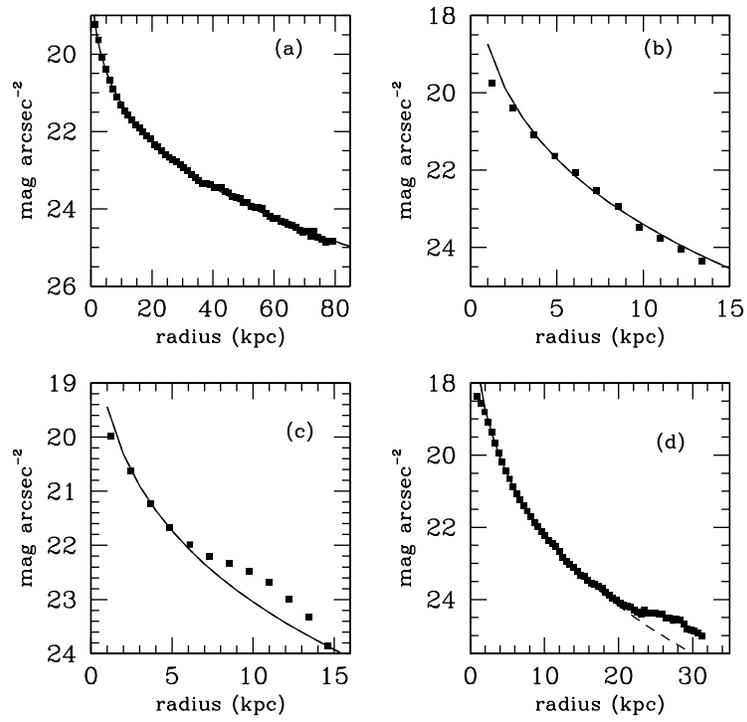}
\caption{Sample galaxies with de Vaucouleurs type profiles.}
\label{fig3}
\end{center}
\end{figure}

\begin{figure}
\begin{center}
\includegraphics[scale=0.6]{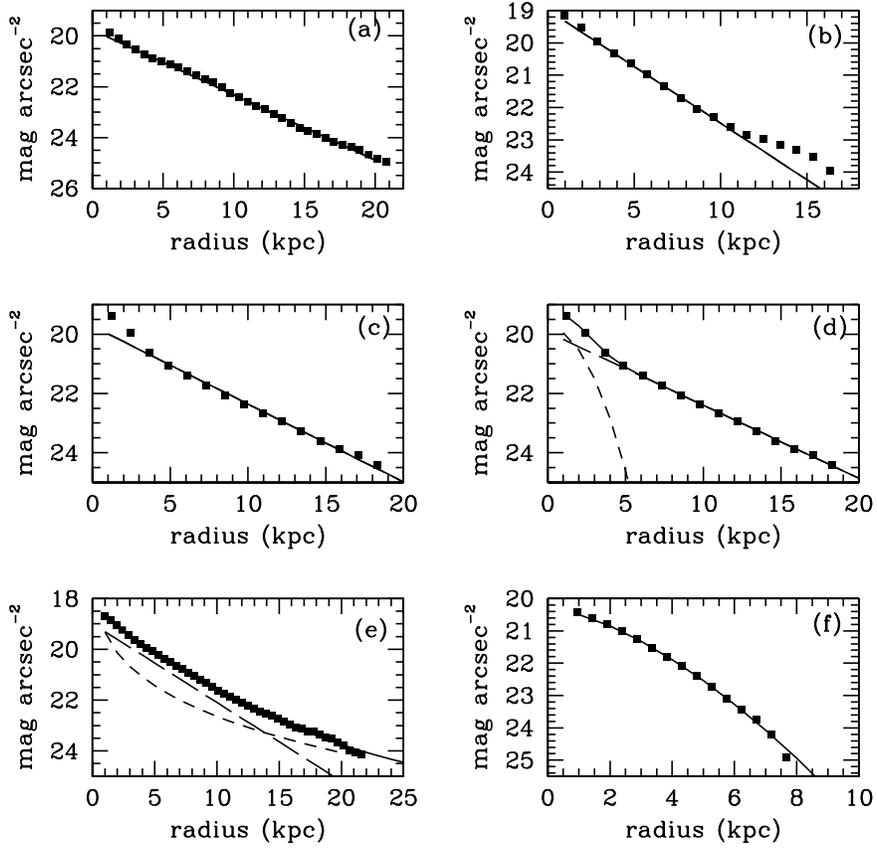}
\caption{Sample galaxies with: (a),(b) and (c): Exponential profiles, (d): exponential (long dashed line) plus gaussian profile (unresolved bulge, short dashed line), (e): bulge (short dashed line) + disk (long dashed line) profile and (f): S\'{e}rsic profile with $n>1$. }
\label{fig4}
\end{center}
\end{figure}

From the total 507 analyzed galaxies, we found that 162 (32\%) have pure $r^{1/4}$ profiles, 168 (33\%) have pure exponential profiles and 93 (18\%) are well fitted by a combination of both. 
We also found 84 (17\%) galaxies with luminosity profiles that could not be fitted with the classical bulge + disk decomposition. For such cases we have only used the S\'ersic law. In addition, we have also fitted the S\'{e}rsic law to the total sample. Figures \ref{fig3} and \ref{fig4} illustrate the cases described above.

\subsection{Photometric scaling relations}

\begin{figure}
\begin{center}
\includegraphics[scale=0.5]{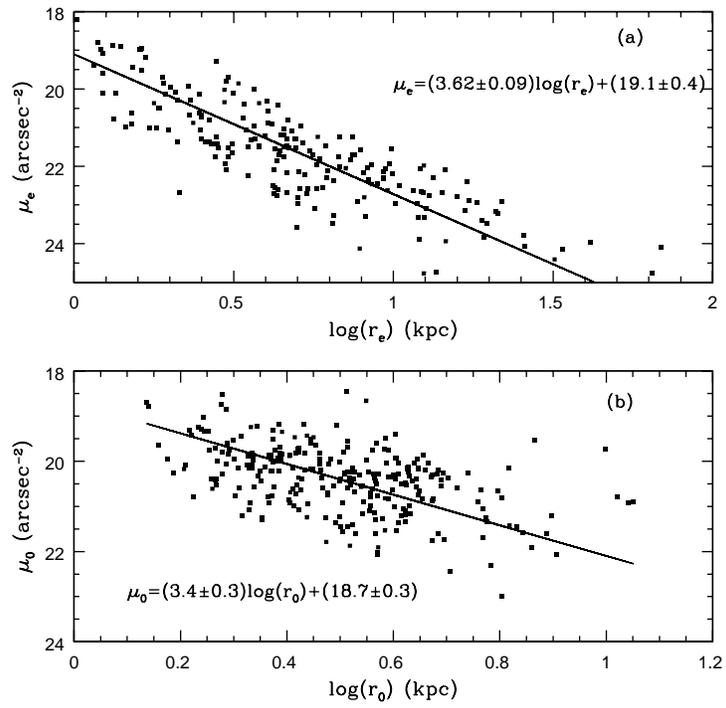}
\caption{Bisector fit for the correlation between the structural parameters for bulges (a) and disks (b).}
\label{fig5}
\end{center}
\end{figure}

The study of correlation between the photometric parameters is a good tool to understand the structure of  galaxies and their evolution. Many scaling relations have been discovered for galaxies, such as the fundamental plane (Djorgovski \& Davis 1987, Dressler et al. 1987) for elliptical galaxies and the Tully-Fisher (1977) relation for spirals. In the absence of spectroscopic measurements of the velocity dispersion, one can study the projection of the fundamental plane which is purely photometric. This plane is defined by the effective radius $r_e$ and the surface brightness $\mu_e$, known as Kormendy relation (Kormendy 1977): $\mu_e=alog(r_e)+b$.  We show this relation in Figure \ref{fig5} (a) and we find a slope value $a =$3.62$\pm$0.09. All slopes and zero points determined in this work correspond to the bisector fit of two ordinary least-squares regressions of $y$ vs $x$ and $x$ vs $y$ as described by Isobe et al. 1990. Our results are in good agreement with those derived by La Barbera et al. (2004) for Abell 2163. Although there are other published results (Ziegler et al. 1999, Andredakis et al. 1995 and M\"{o}llenhoff \& Heidt 2001) they correspond to studies in other photometric bands. We have also studied the $\mu_0-log(r_0)$ relation for disks which is analogous to that of bulges. However, this relation shows a larger scatter than the $\mu_e-log(r_e)$ relation, see Figure \ref{fig5} (b). In this case we find a slope value $a = 3.4\pm0.3$.  

\begin{figure}
\begin{center}
\includegraphics[scale=0.5]{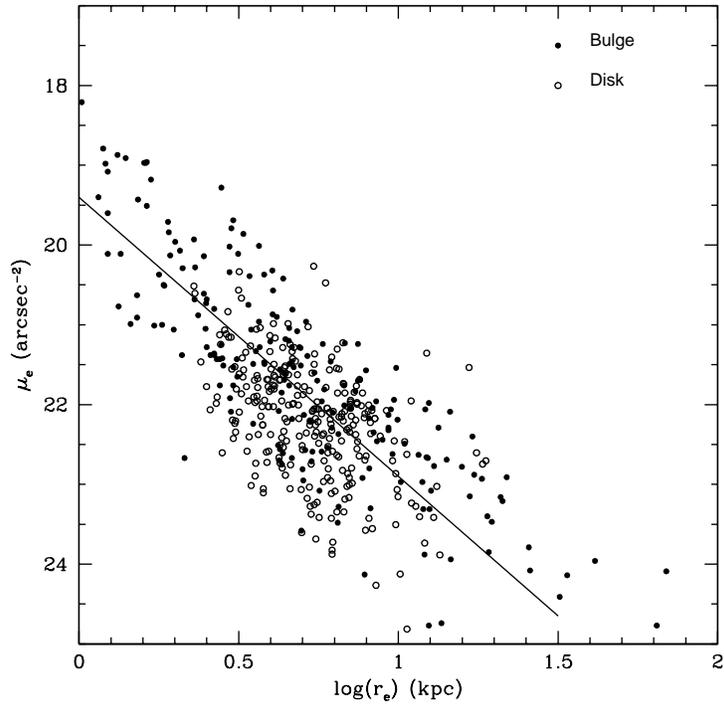}
\caption{Bisector fit for the correlation between the effective surface brightness and the effective radius for bulges and disks.}
\label{fig6}
\end{center}
\end{figure}

In order to compare bulge and disk relations we also determined the effective surface brightness and the effective radius of the disks, as in M\"ollenhoff \& Heidt (2001). Results are plotted in Figure \ref{fig6} where we can observe a similar relation for both subsystems: $\mu_e= (3.5\pm0.2)log(r_e)+(19.4\pm0.4)$. The same relation was previously obtained by M\"ollenhoff \& Heidt (2001) for field spiral galaxies. The results agree with the idea that bulges and disks are 
virialized systems having similar mass-to-light ratio (Bender et al. 1997). 

\begin{figure}
\begin{center}
\includegraphics[scale=0.6]{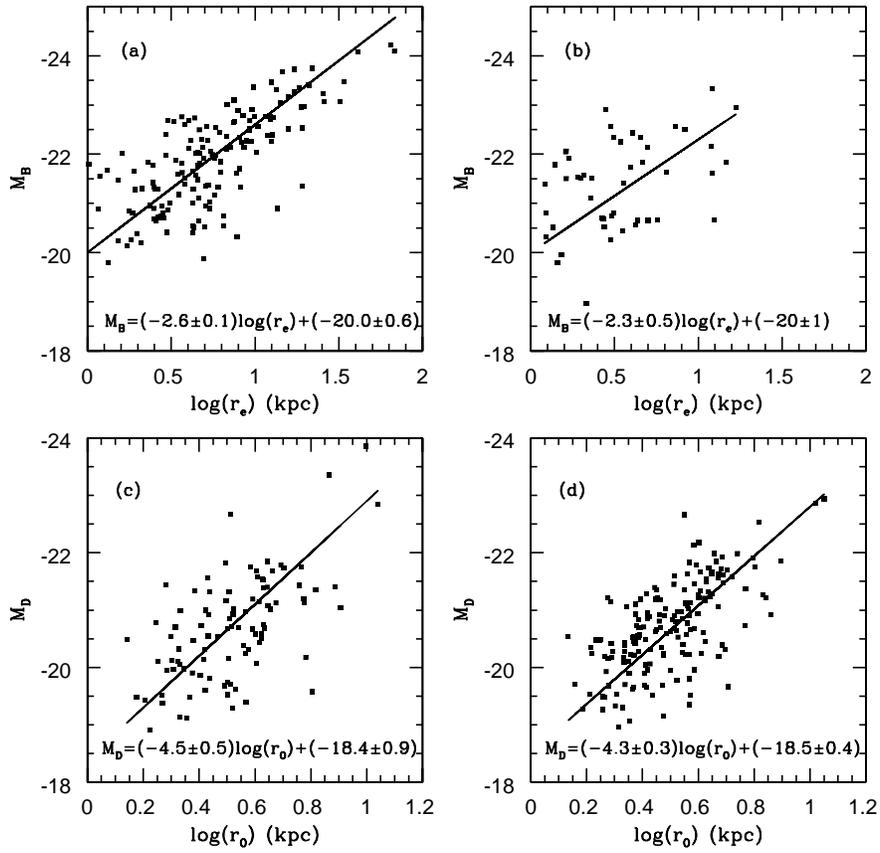}
\caption{Bisector fit for the correlation between the effective radius and the bulge absolute magnitude for $n\le0.4$ (a) and $n>0.4$ (b); and for the correlation between the scale length and disk absolute magnitude for $n\le0.7$ (c) and $n>0.7$ (d).}
\label{fig7}
\end{center}
\end{figure}

We have also analyzed the magnitude-size relation $M_B-log(r_e)$ for bulges (which is closely related to the Kormendy relation), and the $M_D-log (r_0)$ relation for disks. Figure \ref{fig7} shows these relations
where we plot in two different diagrams galaxies with $n\le0.4$ and $n>0.4$. This choice relies on the fact that we observe that luminosity profiles with $n>0.4$ could also be well fitted with the typical bulge + disk function.
In the same way we have considered disks with $n\le0.7$ and $n>0.7$ since those galaxies with $n>0.7$ do not show a bulge component.
However, we found that the relations for each component are independent of the value of $n$ with $M_{B}=(-2.6\pm 0.1)log(r_{e})+(-19.9\pm 0.6)$ for bulges, and $M_{D}=(-4.4\pm 0.2)log(r_{0})+(-18.5\pm 0.4)$ for disks. Our results for disks are in good agreement with those found by Thomas (2002), who uses galaxies from ENACS to study the magnitude-size relations. This author found that the slopes of these relations are not significantly different for different morphologies, so he adopted a single slope (-4.5) for bulges as well as for disks (-4.3). However, in our case bulges show a greater value than that calculated by Thomas (2002).

\begin{figure}
\begin{center}
\includegraphics[scale=0.6]{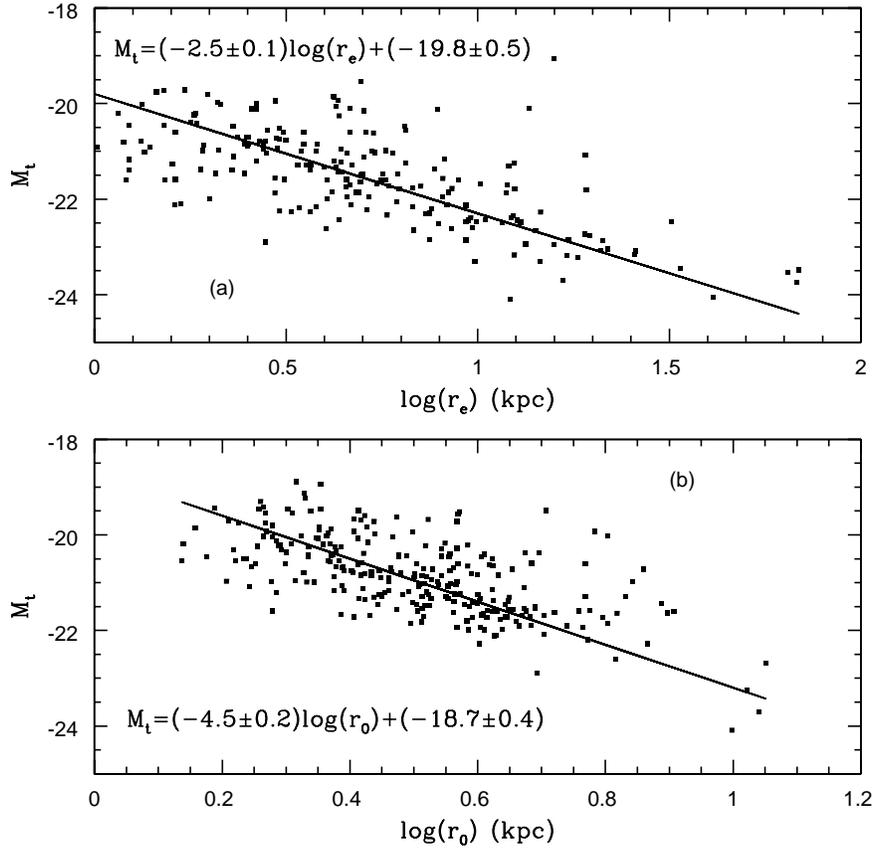}
\caption{Bisector fit for the $M_t$ versus $log(r_e)$ (a) and $M_t$ versus $log(r_0)$ (b) relations. }
\label{fig8}
\end{center}
\end{figure}

We have also studied galaxy total magnitudes and their dependence on bulge and disk properties.
Figures \ref{fig8} (a) and (b) show $M_t$ versus $log(r_e)$ and $M_t$ versus $log(r_0)$ relations. As expected, galaxies with larger bulges and larger disks have higher total luminosity although a large dispersion can be appreciated.

We have described several photometric relations for galaxies of different morphological types. In particular, we found a similar relation between the structural parameters $r_e$, $r_0$, $\mu_e$, $\mu_0$ and the total luminosity of the galaxies and their components. These observed relations contain information about the physical processes that drove their formation and evolution.
Lima Neto, Gerbal \& M\'arquez (1999) found that elliptical galaxies have a quasi-constant specific entropy. On the other hand, M\'arquez et al. (2001) have studied scaling relations between potential energy and mass, and they explain the origin of several observed photometric relations. They found that the theoretical slope of the $L-r_e$ relation (which is related to the $\mu_e-log(r_e)$ relation), depends only on the slope of the scaling relation between the mass and the potential energy of a galaxy. Although this result applies only to elliptical galaxies, it is interesting to note that our sample of disk galaxies obeys
a similar relation. 

\begin{figure}
\begin{center}
\includegraphics[scale=0.5]{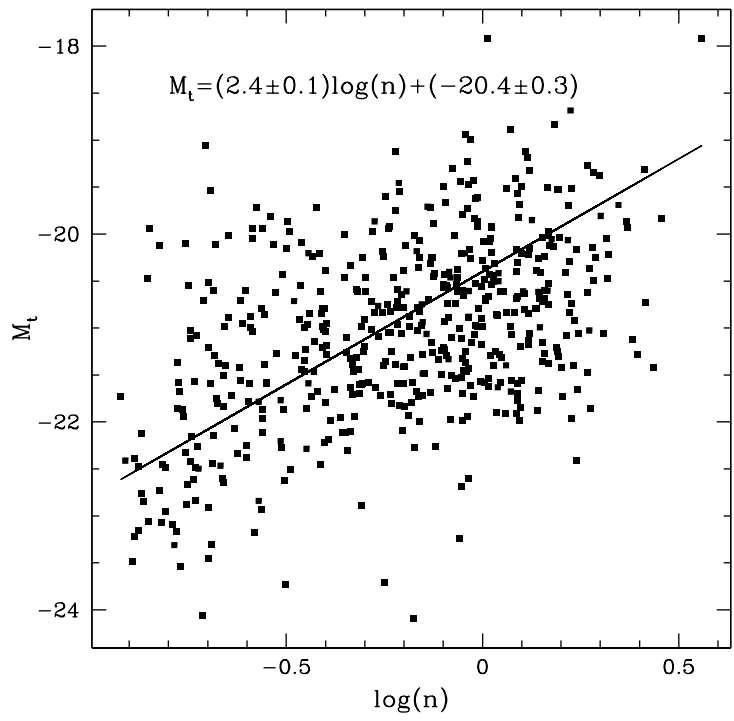}
\caption{Bisector fit corresponding to the correlation between $n$ and the total absolute magnitude.}
\label{fig9}
\end{center}
\end{figure}

We have also investigated the correlation between the shape parameter $n$ and the total luminosity. Using dwarf elliptical galaxies Young \& Currie (1994) have proposed this relation as a distance indicator. They used a small sample of galaxies in the Fornax cluster and found a tight $n-m_t$ relation. We plot the $M_t-log(n)$ relation for the total sample in Figure \ref{fig9} where a clear trend can be appreciated although with a large dispersion that precludes precise distance estimates. However, since our sample includes all type of giant galaxies the observed large dispersion has most likely a different physical origin than the results derived from early-type dwarf galaxies.  

\begin{figure}
\begin{center}
\includegraphics[scale=0.6]{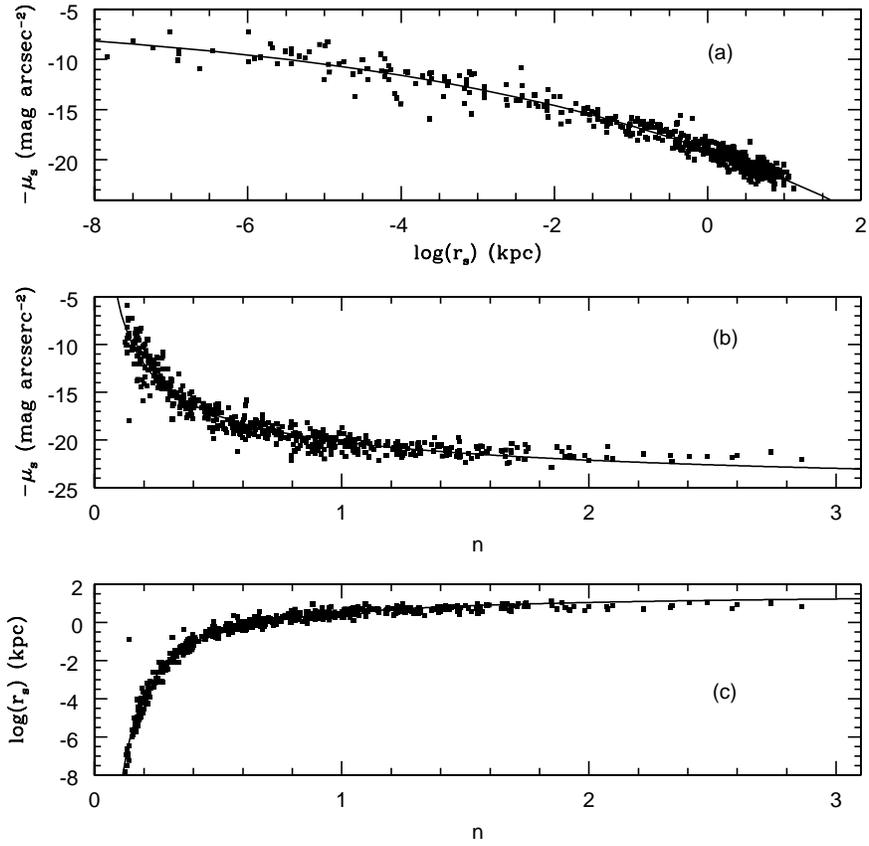}
\caption{(a) $\mu_s$ versus the logarithm of scale radius $r_s$. The solid line is our best fit. (b) $\mu_s$ versus $n$. (c) Logarithm of $r_s$ versus $n$.}
\label{fig10}
\end{center}
\end{figure}

Figure \ref{fig10} shows the correlation between the parameters corresponding to the S\'{e}rsic profile. As we can observe none of these relationships show a linear behavior. For our case the best empirical fitting form for the $\mu_s$ - $r_s$ relation is (panel b):

\begin{equation} \label{s1}
-\mu_s=(-5.2\pm0.6)+(-13.8\pm0.6)exp\Big(\frac{log(r_s)}{-5.2\pm0.3}\Big)
\end{equation}

while for the $\mu_s-n$ relation (see in panel b) is:

\begin{equation} \label{s2}
-\mu_s=(-26.8\pm0.1)+\frac{(6.63\pm0.09)}{n^{0.5}}
\end{equation}

Finally in panel (c) we can observe the $log(r_s)-n$ relation and our best fitting is:

\begin{equation} \label{s3}
log(r_s)=(1.61\pm0.03)+\frac{(-1.11\pm0.01)}{n}
\end{equation}

\begin{figure}
\begin{center}
\includegraphics[scale=0.6]{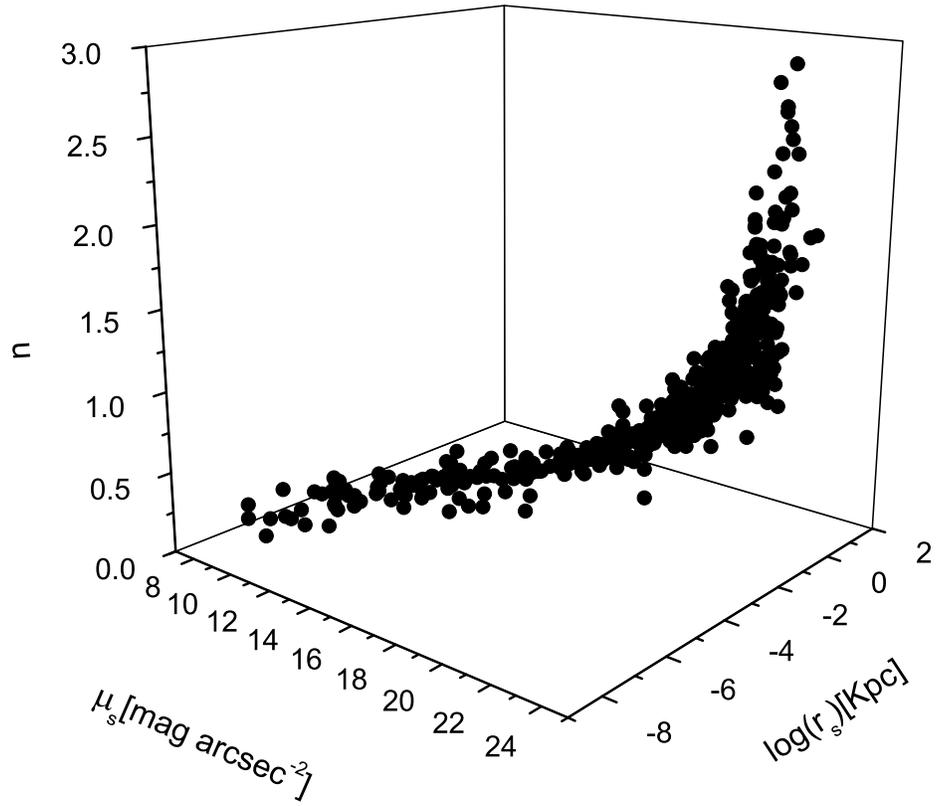}
\caption{3D representation of the correlation among the S\'ersic parameters in the space $[n,log(r_s),\mu_s]$.}
\label{fig11}
\end{center}
\end{figure}

M\'arquez et al. (2001) have demonstrated that the specific entropy and scaling relation between potential energy and mass constrains the distribution of gravitational matter in elliptical galaxies. Moreover, one can define two surfaces related to the physical parameters of the systems that intersect in the $[n,r_s,\mu_s]$ space. From the relation between specific entropy and mass, these authors obtain the \textit{Entropic Surface} expressed in terms of the S\'ersic parameters. In addition, they use the scaling relation between potential energy and mass to obtain the \textit{Energy-Mass Surface}. The intersection line of these two surfaces in  the $[n,r_s,\mu_s]$ space, called the \textit{Entropy-Energy line}, is the only locus along which elliptical galaxies can be found. In other words, elliptical galaxies are indeed a one-parameter family. According to our analysis, not only ellipticals, but all galaxies in the sample lie along this locus. This behavior is shown in Figure \ref{fig11}. We also notice that the upper limit of the observed $r_s$ parameter is $\sim 1 Kpc$, in good agreement with the data and model predictions of M\'arquez et al. (2001).

\section{Conclusions}\label{concl}

We have obtained the luminosity profiles of 507 galaxies in 14 nearby clusters of the SARS survey (Way et al. 2004). We fitted de Vaucouleurs law for the bulge component and an exponential law for the disk component. We also fitted S\'{e}rsic's law for all galaxies of our sample.

We have studied error sources in the photometry like seeing, photon noise and sky cleaning in order to check the uncertainties in the computed parameters. Particularly, sky cleaning, seeing and galaxy overlapping were the main error sources in our case. On the other hand, sky level has a deep influence on the faint end of the luminosity profiles and therefore, on most of the parameters of the luminosity profiles. We found that the error of the shape parameter $n$ is on average 0.08. Measured errors for $\mu_e$, $\mu_0$ and $\mu_s$ were below 0.25 mag arcsec$^{-2}$, while those for $r_e$, $r_0$ and $r_s$ were smaller than 20\%. Finally, we analyzed how seeing affects these  parameters and we determined the optimum radius range  that  minimizes seeing effects on the fitting parameters. For this reason, we have only included in the present analysis those
galaxies with an apparent radius greater than 3$FWHM$ equivalent to an apparent radius $>$ 6$\arcsec$.

The correlation of structural and photometric parameters of bulges and disks show consistency with a unique relation in the $\mu_e - log(r_e)$ plane. As expected, we also found that bulges and disks obey a magnitude-size relation in the sense that large bulges and disks (large $r_e$ values) have high luminosities. The derived $M_D-log(r_0)$ relation is consistent with Thomas (2002), although the $M_B-log(r_e)$ relation is less steep than that obtained  by this author. We find that these magnitude-size relations do not depend on the S\'ersic fit shape parameter $n$ indicating that early-type galaxies and bulges of spirals show a similar behavior.

We investigated the relation between the $n$ parameter and the absolute magnitude finding a weak linear trend. On the other hand, S\'ersic law fitting parameters $n$, $rs$ and $\mu_s$ show a strong correlation in agreement with the theoretical analysis made by M\'arquez et al. (2001). This consistency between our observations and model predictions strongly suggest that galaxies are likely to be understood as a one parameter family.

\section{Acknowledgments}
This work was partially supported by the Consejo de Investigaciones Cient\'{\i}ficas y T\'ecnicas de la Rep\'ublica Argentina, CONICET; SeCyT, UNC, Agencia Nacional de Promoci\'on Cient\'{\i}fica and Agencia C\'ordoba Ciencia, Argentina. L. Infante anf H. Quintana acknowledge partial support from the Centro de Astrof\'{\i}sica FONDAP/CONICYT program.


\clearpage

\clearpage

\renewcommand{\arraystretch}{.6}

\begin{deluxetable}{cccccll}
\tablecaption{Observed clusters.\label{table1}}
\tablewidth{0pc}
\tablehead{
\colhead{Abell Number} &
\colhead{$\alpha_{J2000}$} & 
\colhead{$\delta_{J2000}$} & 
\colhead{Date}&
\colhead{Exposition}&
\colhead{$\sigma$}&
\colhead{$v_r$}\\
&
\colhead{$^h\;\; ^m\;\;\; ^s\;\;\,$} & 
\colhead{$\degr\;\;\; \arcmin\;\;\;\arcsec\;$} & 
&
\colhead{$[frames \& s]$}&
\colhead{[$kms^{-1}$]} &
\colhead{[$kms^{-1}$]}
}
\startdata
118 & 00 55 43.9 & $-$26 24 46 & 17.08.93 & 3 $\times$ 900 & 669 $\pm$ 127 & 34421 $\pm$ 159 \\ 
2734 & 00 11 20.1 & $-$28 52 52 & 18.08.93& 3 $\times$ 900& 784 $\pm$ 124 & 18502 $\pm$ 100 \\
2799 & 00 35 3.00 & $-$39 25 29 & 16.08.93& 3 $\times$ 900& 563 $\pm$ 62  & 19454 $\pm$ 127 \\ 
2800 & 00 37 58.7 & $-$25 05 30 & 17.08.93& 3 $\times$ 900& 335 $\pm$ 64  & 18943 $\pm$ 47 \\
2854 & 01 00 48.7 & $-$50 31 51 & 21.08.93& 1 $\times$ 900& 308 $\pm$ 44  & 18480 $\pm$ 51 \\
2923 & 01 32 18.0 & $-$31 05 36 & 21.08.93& 1 $\times$ 900& 670 $\pm$ 76  & 21420 $\pm$ 135 \\
2933 & 01 40 41.2 & $-$54 33 26 & 21.08.93& 3 $\times$ 900& 759 $\pm$ 72  & 27709 $\pm$ 105\\
3764 & 21 26 1.00 & $-$34 47 39 & 15.08.93& 3 $\times$ 900& 795 $\pm$ 123 & 22714 $\pm$ 110 \\
3809 & 21 49 51.7 & $-$43 52 55 & 17.08.93& 3 $\times$ 900& 560 $\pm$ 67  & 18785 $\pm$ 81 \\
3864 & 22 30 14.4 & $-$52 28 38 & 19.08.93& 3 $\times$ 900& 847 $\pm$ 188 & 30699 $\pm$ 161 \\
3915 & 22 47 37.0 & $-$52 03 09 & 20.08.93& 3 $\times$ 900& 815 $\pm$ 102 & 28925 $\pm$ 105 \\
3921 & 22 49 38.6 & $-$64 23 15 & 18.08.93& 3 $\times$ 900& 788 $\pm$ 111 & 27855 $\pm$ 105 \\
4010 & 23 31 10.3 & $-$36 30 26 & 15.08.93& 3 $\times$ 900& 743 $\pm$ 140 & 28766 $\pm$ 149 \\
4067 & 23 58 48.3 & $-$60 38 39 & 17.08.93& 3 $\times$ 900& 738 $\pm$ 442 & 29643 $\pm$ 181 \\
\enddata
\end{deluxetable}

\begin{deluxetable}{cccc||cccc}
\tablewidth{0pc}
\tablecaption{Obtained photometric parameters for the model galaxies of Fig. 2.\label{table2}}
\tablehead{
\colhead{Galaxy 1}&\colhead{}&\colhead{} &\colhead{} &
\colhead{Galaxy 2} &\colhead{} &\colhead{} & \colhead{}\\
\tableline
\colhead{Fitting region [$\arcsec$]} &
\colhead{$\mu_{s}$ [mag arcsec$^{-2}$]}&
\colhead{$r_{s} $[$\arcsec$]}&
\colhead{$n$}&
\colhead{Fitting region [$\arcsec$]} &
\colhead{$\mu_{s}$ [mag arcsec$^{-2}$]}&
\colhead{$r_{s} $[$\arcsec$]}&
\colhead{$n$}\\
}
\startdata
 Original &19.44 & 3.7 & 0.5 & Original & 20.39 & 2.6 & 1.0 \\
parameters&      &     &     &parameters&       &     &     \\
\tableline
 $r>0.6$ & 19.57 & 4.5 & 0.53 & $r>0.6$ & 20.78 & 3.5 & 1.20 \\
 $r>1.2$ & 19.54 & 4.3 & 0.53 & $r>1.2$ & 20.74 & 3.4 & 1.10 \\
 $r>2.4$ & 19.51 & 4.1 & 0.52 & $r>1.8$ & 20.69 & 3.2 & 1.10 \\
 $r>4.9$ & 19.50 & 4.0 & 0.51 & $r>2.4$ & 20.63 & 3.1 & 1.10 \\
\enddata
\end{deluxetable}

\begin{deluxetable}{cccccccllllllll}
\tabletypesize{\tiny}
\rotate
\tablewidth{0pt}
\tablecaption{\small Structural and photometric parameters: the first column gives the cluster name, column (2) and (3) give the galaxy coordinates 2000.0. Next seven columns give the structural parameters, ie, $m_e$, $m_0$ and $m_s$ expressed in mag arcsec$^{-2}$ and $r_e$, $r_0$ and $r_s$ expressed in Kpc. Column (11) gives the observed absolute magnitude while columns (12) and (13) give the absolute magnitude for the bulge and disk components obtained through profile decomposition. Column (14) gives the absolute magnitude for the unresolved bulges. Finally column (15) gives the sum of bulge and disk luminosities.\label{table3}}
\tablehead{
\colhead{Number} & 
\colhead{$\alpha_{J2000}$} & 
\colhead{$\delta_{J2000}$} & 
\colhead{$\mu_e$}&
\colhead{$r_e$}&
\colhead{$\mu_0$}&
\colhead{$r_0$}&
\colhead{$\mu_s$}&
\colhead{$r_s$}&
\colhead{$n$}&
\colhead{$M_t$}&
\colhead{$M_B$}&
\colhead{$M_D$}&
\colhead{$M_G$}&
\colhead{$M_{ta}$}\\
\colhead{Cluster} & \colhead{[$^h\;\; ^m\;\;\; ^s$]}& \colhead{[$\degr\;\;\; \arcmin\;\;\;\arcsec\;$]}&\colhead{[mag arcsec$^{-2}$]}& \colhead{[Kpc]}&\colhead{[mag arcsec$^{-2}$]}& \colhead{[Kpc]}&\colhead{[mag arcsec$^{-2}$]} &\colhead{[Kpc]} &\colhead{} &\colhead{} & \colhead{}& \colhead{}& \colhead{} &\colhead{} \\
\colhead{(1)} &\colhead{(2)} &\colhead{(3)} &\colhead{(4)} & \colhead{(5)} & \colhead{(6)} & \colhead{(7)} & \colhead{(8)} & \colhead{(9)} & \colhead{(10)} & \colhead{(11)} & \colhead{(12)} & \colhead{(13)} & \colhead{(14)}&\colhead{(15)}\\
}
\startdata
A0118&  00 55 40.15 & -26 11 27.30 &- & - & 20.64 &   3.97 & 20.98 & 0.5220E+01 & 1.21 & -20.83 &  - & -20.93 &  - & -20.93 \\
&  00 55 31.06 & -26 12 44.80 &- & - & - & - & 21.08 & 0.6337E+01 & 1.66 & -20.83 &  - &  - &  - & -20.86 \\
& 00 55 42.37 & -26 12 41.10 &- & - & 20.23 &   3.58 & 20.56 & 0.4655E+01 & 1.18 & -21.07 &  - & -21.12 &  - & -21.12 \\
&  00 55  09.11 & -26 14 28.00 &- & - & 20.24 &   3.26 & 20.74 & 0.4756E+01 & 1.31 & -20.85 &  - & -20.91 &  - & -20.91 \\
&  00 54 35.41 & -26 15 14.20 &21.70 &   5.57 & - & - & 14.40 & 0.5800E-02 & 0.28 & -21.59 & -21.99 &  - &  - & -21.99 \\
&  00 54 59.52 & -26 15 55.40 &21.61 &   4.41 & - & - & 16.11 & 0.4350E-01 & 0.35 & -21.23 & -21.58 &  - &  - & -21.58 \\
&  00 55 15.31 & -26 15 56.30 &- & - & 20.39 &   4.02 & 19.75 & 0.2320E+01 & 0.77 & -21.33 &  - & -21.21 &  - & -21.21 \\
&  00 55 30.17 & -26 15 48.10 &- & - & - & - & 21.37 & 0.6177E+01 & 1.44 & -20.60 &  - &  - &  - & -20.64 \\
&  00 54 52.47 & -26 16 08.70 &- & - & 21.74 &   4.96 & 21.76 & 0.5075E+01 & 1.01 & -20.37 &  - & -20.32 &  - & -20.32 \\
&  00 55 16.32 & -26 16 43.10 &- & - & - & - & 21.23 & 0.1053E+02 & 1.75 & -21.66 &  - &  - &  - & -21.78 \\
&  00 55 13.54 & -26 19 16.90 &22.06 &  12.14 & 19.72 &   9.96 & 18.23 & 0.3596E+01 & 0.67 & -24.09 & -23.33 & -23.85 &  - & -24.37 \\
&  00 55 06.56 & -26 18 54.40 &22.97 &  10.18 & - & - & 17.34 & 0.1305E+00 & 0.38 & -21.68 & -22.04 &  - &  - & -22.04 \\
&  00 54 43.48 & -26 19 12.10 &22.17 &   8.09 & - & - & 14.62 & 0.7250E-02 & 0.28 & -21.96 & -22.34 &  - &  - & -22.34 \\
&  00 55 09.88 & -26 19 33.30 &- & - & 19.87 &   4.52 & 20.20 & 0.5887E+01 & 1.19 & -21.56 &  - & -21.98 &  - & -21.98 \\
&  00 54 43.09 & -26 19 46.00 &- & - & - & - & 20.35 & 0.7830E+01 & 1.69 & -21.97 &  - &  - &  - & -22.04 \\
&  00 55 45.45 & -26 20 05.90 &- & - & - & - &  9.72 & 0.1450E-07 & 0.12 & -22.41 &  - &  - &  - & -23.00 \\
&  00 55 12.87 & -26 19 50.80 &- & - & - & - & 20.52 & 0.9932E+01 & 1.74 & -22.41 &  - &  - &  - & -22.37 \\
&  00 55 43.91 & -26 20 30.20 &- & - & 21.59 &   6.97 & 21.53 & 0.6583E+01 & 0.96 & -20.98 &  - & -21.21 &  - & -21.21 \\
&  00 55 29.18 & -26 20 51.20 &- & - & 20.55 &   4.90 & 20.23 & 0.3654E+01 & 0.86 & -21.54 &  - & -21.48 &  - & -21.48 \\
&  00 54 49.23 & -26 22 16.40 &23.15 &  16.73 & 20.93 &  10.98 & 18.72 & 0.2218E+01 & 0.56 & -23.71 & -22.94 & -22.85 &  - & -23.65 \\
&  00 55 00.81 & -26 21 44.40 &22.69 &  14.17 & - & - & 11.23 & 0.2900E-04 & 0.18 & -22.66 & -23.04 &  - &  - & -23.04 \\
&  00 55 22.20 & -26 31 10.80 &- & - & - & - & 21.72 & 0.6685E+01 & 1.41 & -20.41 &  - &  - &  - & -20.48 \\
&  00 54 51.47 & -26 31 14.50 &- & - & 20.64 &   5.81 & 19.41 & 0.1943E+01 & 0.65 & -21.93 &  - & -21.76 & -19.60 & -21.90 \\
&  00 54 24.80 & -26 30 05.90 &19.96 &   2.00 & 20.04 &   4.07 & 18.43 & 0.1493E+01 & 0.69 & -21.99 & -21.51 & -21.58 &  - & -22.30 \\
&  00 55 09.57 & -26 29 42.30 &- & - & 21.05 &   5.87 & 21.12 & 0.6307E+01 & 1.05 & -21.30 &  - & -21.37 &  - & -21.37 \\
&  00 55 20.26 & -26 30 15.40 &- & - & 20.43 &   3.68 & 19.16 & 0.1131E+01 & 0.62 & -21.05 &  - & -20.97 &  - & -20.97 \\
&  00 55 27.02 & -26 22 04.30 &21.29 &   4.94 & - & - & 17.16 & 0.2755E+00 & 0.46 & -21.63 & -22.15 &  - &  - & -22.15 \\
&  00 54 45.78 & -26 28 33.60 &- & - & - & - & 21.80 & 0.1115E+02 & 2.40 & -21.12 &  - &  - &  - & -21.19 \\
&  00 55 14.34 & -26 27 06.10 &- & - & - & - & 20.82 & 0.6641E+01 & 1.44 & -21.27 &  - &  - &  - & -21.34 \\
&  00 55 09.65 & -26 27 08.50 &- & - & - & - & 21.75 & 0.7032E+01 & 1.46 & -20.60 &  - &  - &  - & -20.52 \\
&  00 55 42.45 & -26 26 11.50 &- & - & 20.48 &   4.19 & 20.94 & 0.6032E+01 & 1.31 & -21.14 &  - & -21.21 &  - & -21.21 \\
&  00 54 58.48 & -26 22 30.80 &22.78 &  15.81 & - & - &  9.79 & 0.1450E-05 & 0.16 & -22.95 & -23.18 &  - &  - & -23.18 \\
&  00 55 28.70 & -26 23 25.70 &- & - & 20.82 &   3.84 & 20.58 & 0.3147E+01 & 0.90 & -20.65 &  - & -20.68 &  - & -20.68 \\
&  00 55 02.72 & -26 22 48.20 &21.65 &   3.13 & 21.34 &   5.97 & 19.68 & 0.1972E+01 & 0.66 & -21.58 & -20.80 & -21.12 &  - & -21.72 \\
&  00 55 18.04 & -26 25 28.20 &20.39 &   3.42 & 21.25 &   5.93 & 16.42 & 0.1595E+00 & 0.41 & -22.19 & -22.25 & -21.20 &  - & -22.60 \\
&  00 55 25.10 & -26 24 59.80 &- & - & 20.16 &   4.44 & 20.58 & 0.6220E+01 & 1.26 & -21.56 &  - & -21.66 &  - & -21.66 \\
&  00 55 34.15 & -26 24 33.90 &- & - & 19.75 &   3.64 & 19.58 & 0.3175E+01 & 0.93 & -21.62 &  - & -21.63 &  - & -21.63 \\
&  00 54 36.38 & -26 23 56.70 &23.08 &  12.66 & - & - &  7.26 & 0.1015E-05 & 0.15 & -22.45 & -22.40 &  - &  - & -22.40 \\
&  00 55 29.91 & -26 23 46.80 &- & - & 21.21 &   7.89 & 21.32 & 0.8816E+01 & 1.09 & -21.63 &  - & -21.85 &  - & -21.85 \\
&  00 54 53.62 & -26 22 52.70 &22.97 &  11.89 & - & - & 11.26 & 0.1160E-04 & 0.17 & -21.87 & -22.37 &  - &  - & -22.37 \\
&  00 55 28.36 & -26 23 16.10 &21.53 &   4.70 & - & - & 15.52 & 0.2900E-01 & 0.33 & -21.29 & -21.80 &  - &  - & -21.80 \\
&  00 55 27.35 & -26 23 13.00 &- & - & 20.18 &   4.44 & 20.20 & 0.4524E+01 & 1.01 & -21.58 &  - & -21.63 &  - & -21.63 \\
&  00 54 47.09 & -26 30 00.10 &23.40 &  19.00 & - & - & 10.22 & 0.1015E-05 & 0.15 & -22.73 & -22.97 &  - &  - & -22.97 \\
\enddata
\tablecomments{The complete version of this table is in the electronic edition of the Journal. The printed edition contains only a sample.}
\end{deluxetable}

\end{document}